\documentclass[12pt,a4paper]{JHEP3}
\usepackage{setspace,epsfig}
\def\gf{\hat{G}^{4}}
\def\gt{\hat{G}^{2}}
\def\hg{\hat{G}}
\def\sg{\hat{G}^{*}}
\def\gl{\hg^{4}\lambda^{16}}
\def\els{\bar{\epsilon}\lambda^{*}}
\def\lam{\lambda^{16}}
\def\dlf{\det{e} \lambda^{15}}
\def\dls{\det{e} \lambda^{16}}
\def\ga{\gamma}
\def\sc{\psi^{*}}
\def\lst{\lambda^{*}}
\def\zerd{\delta^{(0)}}
\def\lof{{L}_{1}^{(5)}}
\def\ltf{{L}_{2}^{(5)}}
\def\lthf{{ L}_{3}^{(5)}}
\def\loz{{L}_{1}^{(0)}}
\def\ltz{{L}_{2}^{(0)}}

\def\bee{\begin{equation}}
\def\ene{\end{equation}}
\def\bea{\begin{eqnarray}}
\def\ena{\end{eqnarray}}
\def\fof{f_{2}^{(14,-14)}}
\def\fot{\tilde{f}_{2}^{(13,-13)}}
\def\fon{f_{2}^{(13,-13)}}
\def\df{{\rm{\bar{D}}}_{-14}}
\def\dt{{\rm D}_{13}}
\def\rrr{\rho_{1}\rho_{2}\rho_{3}}

\def\rfive{\rho_{1}\cdots\rho_{5}}
\def\mnl{\mu\nu\lambda}
\def\la{\lambda}
\def\doz{\delta_{1}^{(0)}}
\def\dtz{\delta_{2}^{(0)}}
\def\dof{\delta_{1}^{(5)}}

\def\tb{\tau_{2}}
\def\bt{\bar{\tau}}
\def\t{\tau}
\def\ptb{\frac{\partial}{\partial{\bar{\t}}}}
\def\pt{\frac{\partial}{\partial{{\t}}}}
\def\nn{\nonumber}
\def\lcd{\lambda^{14}}
\def\gmnru{\gamma^{\mu\nu\rho}}
\def\gmnrd{\gamma_{\mu\nu\rho}}
\def\gz{\gamma^{0}}
\def\es{\epsilon^{*}}
\def\e{\epsilon}
\def\s{\psi}
\def\ega{\bar{\epsilon^{*}}\gamma^{\mu}\sc_{\mu}}
\def\dep{\delta_{\epsilon_{1}}}
\def\deps{\delta_{\epsilon_{2}^{*}}}
\def\epepsg{\bar{\epsilon_{2}}\ga^{\sigma}\epsilon_{1}(\ga_{\sigma})_{ba}}
\def\lb{\bar{\lambda}}
\def\ftwe{f^{(12,-12)}(\t,\bt)}
\def\fele{f^{(11,-11)}(\t,\bt)}
\def\del{{\rm D}_{11}}
\def\dtw{\bar{\rm D}_{-12}}
\def\Dtw{\nabla^{2}_{(-)12}}
\def\Dft{\nabla^{2}_{(-)14}}
\def\hw{\hat{w}}
\def\cD{{\cal D}}
\def\rD{{\rm D}}
\def\brD{\bar{\rm D}}
\def\p{\partial}
\def\om{\mu_{1}}
\def\tm{\mu_{2}}
\def\thm{\mu_{3}}
\def\fm{\mu_{4}}
\def\on{\nu_{1}}
\def\tn{\nu_{2}}
\def\thn{\nu_{3}}
\def\fn{\nu_{4}}
\def\bz{\bar{z}}
\def\tr{\rm tr}
\def\lan{\langle}
\def\ran{\rangle}
\def\bp{\bar{\partial}}
\def\frt{\frac{1}{t}}

\def\a#1{a_{#1}}
\def\b#1{b_{#1}}
\def\m#1{\mu_{#1}}
\def\n#1{\nu_{#1}}
\def\r#1{\rho_{#1}}
\def\bl{\bar{\lambda}}
\def\ff{(-6i\psi\psi)^{4}}
\def\fs{(-6i\psi\psi)^{2}}
\def\ts{\bar{\theta}^{*}}

\title{ The $\hat{G}^{4}\lambda^{16}$ Term in IIB Supergravity}
\author{Aninda Sinha\\Department of Applied Mathematics and Theoretical Physics,\\ Wilberforce Road,\\
Cambridge CB3 0WA, UK
\\Email:
\email{A.Sinha@damtp.cam.ac.uk}}
\abstract{The supersymmetry constraints on the $\gl$ term in the
effective action of type IIB superstring theory are studied in order to
determine the dependence of its coefficient on the complex scalar
field, $\tau$. The resulting expression is consistent with the
$SL(2,{\Bbb Z})$ invariant conjectures in the literature.}

\begin{document}
\onehalfspacing
\section{Introduction}
Chiral $N=2$, $D=10$ supergravity\cite{sch,shwest} is the low energy limit of
type IIB string theory. Higher derivative terms in the low energy
limit can be generated by considering scattering amplitudes
in string perturbation theory. This determines the terms proportional
to $e^{-2\phi}$, where $\phi$ is the dilaton. However, determining the exact
dependence of these terms on the scalar fields is more challenging. In
principle, perturbative contributions can be determined for higher
genus string loop calculations, but there is no direct way of
determining the non-perturbative contributions. The exact action must be invariant under
$SL(2,{\Bbb Z})$ which means that the scalar field dependence is encoded in modular forms,
depending on $\tau=C^{0}+ie^{-\phi}$, where $C^{0}$ is the Ramond-Ramond
scalar.

The constraints imposed by supersymmetry are very powerful.
In \cite{mbgss,Green:1999qt}, it was shown how to use supersymmetry to
compute the coefficient for the 16-dilatino term which appears at
eight-derivative order, i.e., at $\alpha'^{3}$ relative to the tree-level. This term has been analyzed in \cite{mbggut}. The coefficient turns out to satisfy an
eigen-value equation for the laplacian on the fundamental domain of
$SL(2,\Bbb{Z})$. The solution for such an equation is a generalized Eisenstein
series \cite{mbggut2,Green:1999qt,Obers:1999es}. This series has an expansion which encodes tree-level and
higher genus information along with an infinite series of D-instanton
contributions. The 16-dilatino term is related by linearized
supersymmetry to the $C^{4}$ term, where $C$ denotes the Weyl tensor and
$C^{4}$ symbolizes the contraction of four of these. The coefficient of this term is a function
of $\tau,\bar{\tau}$ which has been studied through different consistent arguments\cite{Green:1999pv,Green:1999pu,Peeters:2001ub,Peeters:2000sr,Chalmers:2000zg},
which serve as a powerful countercheck for the validity of the
calculation. The coefficient implies that the $C^{4}$
term gets only tree-level, one-loop and a series of D-instanton contributions. The
validity of this powerful prediction has been checked by explicit
two-loop calculation for the four-graviton in \cite{zhu,iengo}. It was shown
that there is no genus-2 contribution. Furthermore, a
non-renormalization theorem was proved in \cite{berk1} which showed
$C^{4}$ cannot receive perturbative contributions
beyond tree-level and one-loop. 

There have been other ways to infer the coefficients of
higher-derivative terms. In \cite{berk}, an all genus conjecture for
terms like $C^{4}\hat{G}^{4g-4}$ in type IIB in ten dimensions was made, where $\hat{G}$ stands for the supercovariant
antisymmetric three-form field strength and $g$ is the genus. The
argument was motivated by $N=4$ topological string theory \cite{berk2}. Strong
evidence was presented that the coefficient of these terms are higher
order Eisenstein series. All these Eisenstein series have the generic
feature of representing tree-level and genus-$g$ contributions as well
as a series of D-instanton contributions. It should be possible to
prove these conjectures by using the supersymmetric methods of \cite{mbgss}.

Since the $C^{4}$ interaction is related by superspace arguments to the
16-dilatino interaction, it is expected that there will be
$\gl$ term in the action, $\lambda$ being the dilatino, at the
twelve-derivative or order $\alpha'^{5}$ 
relative to the tree-level terms.

In \cite{mbgss}, motivated by \cite{berk,russo}, a conjecture was made for higher derivative extension of the IIB effective action. It reads

\bea
(\alpha')^{4}\sum_{g,\hat{g}=1}^{\infty}\sum_{p=2-2g}^{2g-2}&&(\alpha')^{2g+2\hat{g}-1}\int d^{10}x\,\det{e} F_{5}^{4\hat{g}-4}\hat{G}^{2g-2+p}\hat{G}^{*2g-2-p}\\ \nn
&&\left(f_{g+\hat{g}-1}^{(p,-p)}(\tau,\bar{\tau})C^{4}+\cdots+f_{g+\hat{g}-1}^{(12+p,-12-p)}(\tau,\bar{\tau})\lambda^{16}\right), \label{conj}
\ena
where $F_{5}$ is the self-dual 5-form field strength and $\hat{G}$, the supercovariant version of
the field strength $G$ is 
\bee
\hg_{\mu\nu\rho}=G_{\mu\nu\rho}-3\bar{\psi}_{[\mu}\gamma_{\nu\rho]}\lambda-6i\bar{\psi^{*}}_{[\mu}\gamma_{\nu}\psi_{\rho]},
\ene
where $\psi_{\mu}$ is the gravitino.
The modular forms $f_{g}^{(q,-q)}$ are expected to be given by the generalized Eisenstein series
\bee
f_{g}^{(q,-q)}=\sum_{(m,n)\neq(0,0)}{\tau_{2}^{g+\frac{1}{2}}\over (m+n\tau)^{g+\frac{1}{2}+q}(m+n\bar{\tau})^{g+\frac{1}{2}-q}}.
\ene
For $q=0$ these functions are proportional to $E_{g+\frac{1}{2}}$ where $E_{s}$ is defined in equation (B.12). For $g=2,\hat{g}=1,p=2$ there is evidently a $\det{e}\gl$ term in the integrand. The coefficient of this term is conjectured to be $f_{2}^{(14,-14)}$. In \cite{mbgss}, a schematic method
of obtaining the coefficient of this term was presented using
supersymmetry arguments. However, the calculation was not completed
and though it seemed plausible, the fact that the coefficient of such
a term is a generalized Eisenstein series, was not proved. Such a proof will give
further evidence for the conjectures in the literature for the
ten-dimensional effective action. Implications using the AdS/CFT
correspondence for this term are also currently being investigated \cite{greenkovacs}.

In this paper, we construct a proof using supersymmetry that the
coefficient of the $G^{4}\lambda^{16}$ term is the expected modular form derived from the Eisenstein series, $E_{5/2}$.

The paper is organized as follows. In section 2, we outline the method
used in \cite{mbgss} to obtain the coefficient of the 16-dilatino
term. In section 3, we determine the
coefficient for $\gl$.In section 4, we discuss the tensor structure for
$\gl$. Two appendices have
been included which summarize various identities and supersymmetry
transformations required in the paper.

\section{Review of $\lam$ term at order $\alpha'^{3}$}

In this section we briefly summarize the supersymmetry calculation
at order $\alpha'^{3}$ as done in \cite{mbgss}. The notation is made clear in appendix A. 
There is no off-shell superspace formulation
for the theory as a result of which an action with manifest
supersymmetry cannot be written down. It is possible to write on-shell
superfields \cite{howewest} and use them to write manifestly supersymmetric equations
of motion. In what follows the lagrangian will just be a shorthand for the equations of motion. The low-energy effective action can be
written as 
\bee
S=\int d^{10}x\sqrt{g}L\sim
\frac{1}{\alpha'^{4}}\left[S^{(0)}+\alpha'^{3}S^{(3)}+\alpha'^{4}S^{(4)}+\alpha'^{5}S^{(5)}+...\right],
\ene
where $L$ is the lagrangian and there are no contributions at $\alpha'$ and $\alpha'^{2}$ order. The
supersymmetry transformation $\delta$ can also be expanded in powers
of $\alpha'$.
\bee
\delta\sim
\delta^{(0)}+\alpha'^{3}\delta^{(3)}+\alpha'^{4}\delta^{(4)}+\alpha'^{5}\delta^{(5)}+....
\ene
The Noether method of constructing supersymmetric actions demands that
the supersymmetry transformations close on using the equations of
motion. This will yield supersymmetry constraints of the form

\bee
\delta^{(0)}S^{(0)}=\delta^{(0)}S^{(3)}+\delta^{(3)}S^{(0)}=\delta^{(0)}S^{5}+\delta^{(5)}S^{(0)}=\cdots=0 \label{susy}.
\ene
The
calculation proceeds as follows. Two specific terms in the effective
lagrangian are selected
which do not mix under supersymmetry with any other terms at this order. These are:

\bee
L_{1}^{(3)}=\det
e\left(\ftwe\lam+\fele(\la^{15}\ga^{\mu}\sc_{\mu})\right). \label{al}
\ene
The normalization of the terms have been changed slightly from that in
\cite{mbgss} for our
convenience. In type IIB supergravity, we define two supersymmetry parameters
$\epsilon$ and $\epsilon^{*}$. The lowest
order $\epsilon$ supersymmetry transformations of $L_{1}^{(3)}$
contain a term proportional to $\dls
\sc_{\mu}\e$ with a coefficient that has to vanish for the action to
be supersymmetric, leading to the condition

\bee
\del\fele=4 \ftwe, \label{sub1}
\ene
where the notation $\del$ is explained in Appendix A.
The $\epsilon^{*}$ variation of (\ref{al}) gives a term of the form
$\dls\lst\e^{*}$. In this case, equation (\ref{susy}) can only be
satisfied if account is taken of a term from the
lowest order IIB lagrangian,

\bee
\loz=\frac{1}{256} \det e \bar{\lambda}\ga^{\rrr}\lambda^{*}
\bar{\lambda^{*}}\ga_{\rrr}\lambda .
\ene
A modification to the $\epsilon^{*}$ supersymmetry
transformation of $\lst$, of the form

\bee
\delta_{1}^{(3)} \lst_{a}= g(\t,\bt)\gf(\lcd)_{cd}(\gmnru\gz)_{dc}(\gmnrd\es)_{a},
\ene
where $g(\tau,\bar{\tau})$ is an unknown function, acting on
$L_{1}^{(0)}$ leads to
\bee
2\dtw\ftwe+15\fele-3\cdot 360ig=0 . \label{sub2}
\ene
Finally, we obtain a constraint by demanding the closure of the
supersymmetry algebra on  $\lst$. This gives rise to the relation:

\bee
-192i\del g=\ftwe \label{sub3}
\ene
Combining (\ref{sub1}),(\ref{sub2}),(\ref{sub3}) gives

\bee
\Dtw\ftwe=\left(-132+\frac{3}{4}\right)\ftwe .
\ene
This is exactly what we expect as shown in equation (\ref{al3}) in the
appendix.

\section{$\gl$ term at order $\alpha^{'5}$}
By considering the generalization of the preceding argument to the 
$\alpha'^{5}$ term $\gl$, one can show that its coefficient is a
modular form. 
 The
term involving gravitinos in $\gl$ is schematically
$(\psi\psi)^{4}\lambda^{16}$. Since $\lambda^{16}$ forms a Lorentz
singlet, $\hg^{4}$ will also be a Lorentz singlet. In principle, there
are three independent ways to form a singlet using four $\hg$s,
which are diagrammatically represented in figure 1.

\vskip 1.5cm  

\FIGURE{\epsfig{file=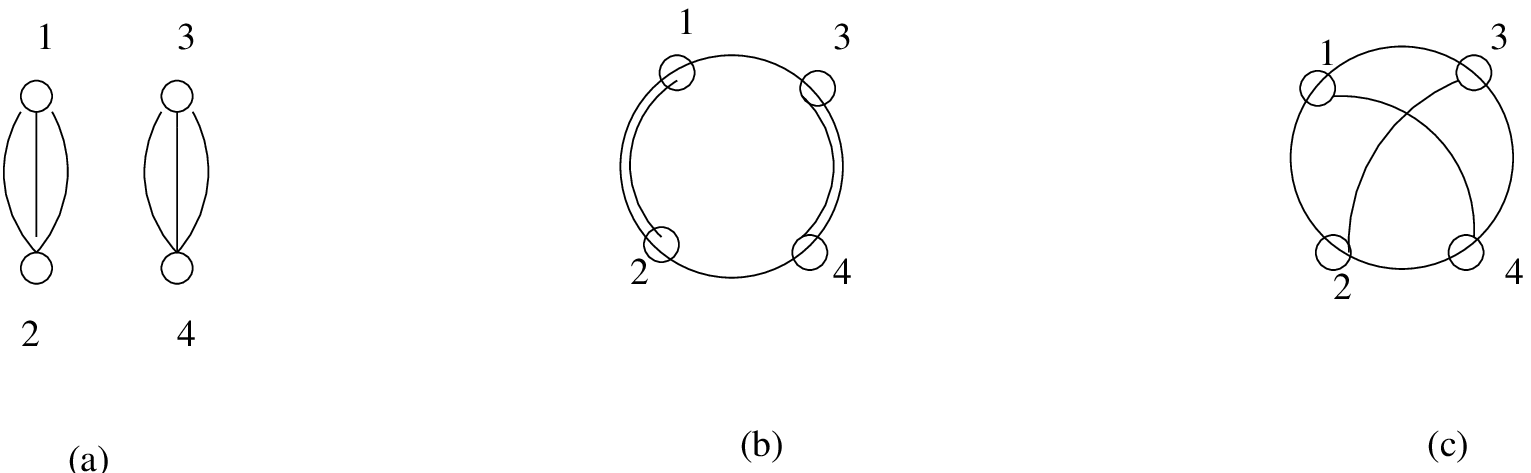, height=4cm, width=10cm}
\caption{Diagrammatic representation of the various contractions. (a)
represents structure $T_{3}$ in the paper, (b) represents $T_{1}$ and
(c) represents $T_{2}$. The lines indicate contractions and the small
circles represent $\hat{G}$.}}

In terms of space-time indices, the contractions are given by
\bea
\hg^{4}= a\,
\hg_{\m1\n1\r1}\hg^{\m1\m2\r1}\hg_{\m3}^{\,\,\,\,\n1\n2}\hg_{\m2\n2}^{\,\,\,\,\,\,\,\,\,\,\m3}&+&b\,
\hg_{\m1\n1\r1}\hg^{\m1\m3\n2}\hg_{\m3\m4}^{\,\,\,\,\,\,\,\,\,\,\,\,\r1}\hg^{\m4\n1}_{\,\,\,\,\,\,\,\,\,\,\n2}+
\nn\\&&c\,\hg_{\m1\n1\r1}\hg^{\m1\n1\r1}\hg_{\m2\n2\r2}\hg^{\m2\n2\r2} ,\label{tensor}
\ena
where $a,b,c$ are undetermined coefficients which are assumed to be
non-zero. The following argument will not yield the values of
$a,b,c$. To be very specific, the term proportional to
$\hg_{\m1\n1\r1}\hg^{\m1\n1\r1}\hg_{\m2\n2\r2}\hg^{\m2\n2\r2}\sim \hg^{2}\hg^{2}$ will be
considered, though it is easy to generalize the argument to the other
two cases.
The piece of $\hg_{\mu\nu\rho}$ involving gravitino bilinears will be written using the
shorthand notation $\ff$.
 
Following \cite{mbgss} we will now select three terms with the
appropriate dimensions contributing to $S^{(5)}$ that will mix with
each other, but with no other terms, under supersymmetry. These are
\bea
\lof &=& \dls\gf \fof \\ \label{L15}
\ltf &=& \dlf \ga^{\mu}\sc_{\mu}\gf \fon \\ \label{L25}
\lthf &=& \dls \gt \hg_{\rrr} \hg^{* \rrr}\fot .  \label{L2t5}
\ena
The $\epsilon^{*}$ supersymmetry variation proportional to
{$\dls\els\ff$} gives
\bea
\doz\lof&=&-2\dls (\els)\ff(\tb\pt-7i)\fof \\ \nn
&=&-2i \dls\els\ff\df\fof ,
\ena
and
\bea
\doz\ltf&=&\dlf\delta(\ga^{\mu}\sc_{\mu})\ff\fon \\ \nn
&=&-15 i \dls\els\ff\fot .
\ena
$\sg$ in $\lthf$ has a term of the form 
$\psi_{\mu} \lambda^{*}$. Taking into account the fact that the
$\epsilon^{*}$ supersymmetry variation of
$\psi_{\mu}$, as given in appendix A,  has a
$\hg$ piece and the $\epsilon^{*}$ variation of $\lambda^{*}$ has a
$\psi_{\mu}\lambda^{*}$ piece, the following equation is obtained

\bee
\doz \lthf=-\frac{9}{16}\fot\dls\ff\els .
\ene
In addition we now consider the $O(\alpha'^{5})$ supersymmetry transformations acting on the
following two terms from the classical action,

\bea
\loz&=&\frac{1}{256} \det e \bar{\lambda}\ga^{\rrr}\lambda^{*}
\bar{\lambda^{*}}\ga_{\rrr}\lambda \\
\ltz&=&-\frac{1}{8}\bar{\sc_{\mu}}\ga_{\nu\rho}\lambda^{*}\hg^{\mu\nu\rho} ,
\ena
where $\loz$ is the same interaction considered in the previous
section and $\ltz$ can be read off from equation (4.12) of
\cite{sch}. The modified $\epsilon^{*}$ supersymmetry
transformations at order $\alpha'^{5}$ are

\bea
\dof \lst_{a}&=& g_{1}(\t,\bt)\gf (\lcd)_{cd}(\gmnru\gz)_{dc}(\gmnrd\es)_{a} \\
\label{lstar}
\dof\s_{\mu_{a}}&=&g_{2}(\t,\bt)\lam\gt\hg_{\rrr}(\ga^{\rrr}\ga_{\mu})_{ab}\es_{b} ,
\label{psi}
\ena
where $g_{1}$ and $g_{2}$ are unknown functions of $\tau$ and $\bar{\tau}$. These transformations
acting on $L_{1}^{(0)}$ and $L_{2}^{(0)}$ give
\bea
\dof\loz&=&-3.360 \det e\,g_{1}\gl\els \\
\dof\ltz&=&-\frac{3}{4}\det e\,g_{2}\gl\els .
\ena
In order to satisfy the constraint ($\ref{susy}$), the following equation is obtained 
\bee
2\df\fof+15\fon-\frac{9i}{16}\fot-3\cdot360i\,g_{1}-\frac{3}{4}i\,g_{2}=0 .
\label{main1}
\ene
Now we consider supersymmetry variations of the form $(\dls \ff) \ega$.
The term  $\lthf$ doesn't mix since $\sg$ has either a
$\sc_{\mu}\sc_{\nu}$ piece or a $\psi_{\mu}\lst$ piece, neither of which
yield $\sc\psi\psi$ under supersymmetry variation.
Variation of $\lof$ yields, 
\bea
\dtz\lof&=(\dtz\det e)\lam\ff\fof+\det e\,(\dtz\lam)\ff\fof \\ \nn
&+\det e\lam(\dtz\ff)\fof .
\ena
Here the first two terms are given by equation (3.4) in
\cite{mbgss}. Together they give  $(2i\dt\fon-8i\fof)\det
e\,\ega\lam$. The last term can be written as:

\bee
\det e\lam(\dtz\ff)\fof=(-\frac{7}{4}-\frac{5}{4})i\dls\ff\ega ,
\ene
where the first term in the bracket on the right comes from the
$\hat{F}_{5}$ in the variation of $\psi_{\mu}$
and the second comes from the supercovariant derivative acting on
$\epsilon$. The supercovariant derivative $D_{\mu}$ has a
piece that depends on the gravitino bilinear \cite{sch} which has
been taken into account. Thus,
\bee
\dtz\lof=(2\dt\fon -(8+\frac{7}{4}+\frac{5}{4})\fof)i\dls\ff\ega ,
\ene

and the action is supersymmetric if 
\bee
2\dt\fon -(8+\frac{7}{4}+\frac{5}{4})\fof=0 . \label{main2}
\ene

Further constraints are imposed by demanding the closure of the supersymmetry algebra. In particular $[\dep,\deps]\lst$ and $[\dep,\deps]\psi_{\mu}$ are considered. In a manner similar to deriving (3.17) of \cite{mbgss} we get

\bea
[\dep^{0},\deps^{5}]\lst_{a}&=&-192i\dt
g_{1}\la^{15}_{b}\left[\frac{3}{8}\epepsg\right]\\ \nn
&+&g_{1}(\dep^{0}\gf)(\lcd)_{cd}(\gmnru\gz)_{dc}(\gmnrd\es_{2})_{a}+\cdots ,
\ena
where the ellipsis indicate terms that are not needed for the analysis.
To evaluate the last term note that $\delta \hg\sim \delta(\psi\psi)+\delta(\sc\la)$. $\delta \sc$ will have a $\hg^{*}$ piece and this will relate $g_{1}$ to $\fot$. The result is

\bee
g_{1}(\dep^{0}\gf)(\lcd)_{cd}(\gmnru\gz)_{dc}(\gmnrd\es_{2})_{a}=-108g_{1}\gt(\hg^{\mnl}\sg_{\mnl})\la^{15}_{b}(\frac{3}{8}\epepsg)+\cdots .
\ene

Thus the commutator of $\delta_{\epsilon_{1}}$ and
$\delta_{\epsilon_{2}^{*}}$ acting on $\lambda^{*}$ is

\bea
[\delta_{\epsilon_{1}},\delta_{\epsilon_{2}^{*}}]\lambda_{a}^{*}&=&\bar{\epsilon_{2}}\gamma^{\mu}\epsilon_{1}\rD_{\mu}\lambda_{a}^{*}+\frac{3}{8}\epepsg\{-i(\gamma^{\mu}\rD_{\mu}\lambda^{*})_{b}\\ \nn
&+&\alpha'^{5}(-192i\dt
g_{1}\la^{15}_{b}-108g_{1}\gt(\hg^{\mnl}\sg_{\mnl})\la^{15}_{b})\}+\cdots,
\ena
where the first two terms on the right-hand-side come from the commutator of the
lowest order supersymmetry transformations\cite{sch,mbgss}.
In order to close the
algebra, the equations of motion have to be used. These give

\bea
-192i \dt g_{1}&=&\fof \\
-108 g_{1}&=&\fot . \label{main3}
\ena

Terms proportional to
$(\epsilon_{1}\la^{15}\gf\epsilon^{*}_{2})_{\mu}$ in
$[\dep,\deps]\s_{\mu}$ are now considered. Using various Fierz transformations
and gamma product expansions we find

\bee
\dep^{0}\deps^{5}(\psi_{\mu})_{c}=\frac{i}{64}g_{2}\la^{15}_{a}\gf(\ga_{\mu}\ga^{\sigma})_{ac}(\bar{\epsilon_{2}}\ga_{\sigma}\epsilon_{1})+\cdots .
\ene

In addition, $\deps^{5}\dep^{0}\psi_{\mu}$ contributions proportional
to $\la^{15}_{a}\gf(\ga_{\mu}\ga^{\sigma})_{ac}(\bar{\epsilon_{2}}\ga_{\sigma}\epsilon_{1})$
also need to be taken into account. $\delta^{(0)} \psi_{\mu}$ has
terms proportional to $\la\lst$ (See equation (A.6) in the appendix). These come from the $\hat{F}_{5}$ piece as well as the other $\la\lst$ terms in $\delta \psi_{\mu}$.
The relevant terms in the supersymmetry transformation for $\psi_{\mu}$ proportional to $\epsilon$ are
\bea \label{g1}
\delta^{0}_{\epsilon_{1}}\s_{\mu}&=-\frac{i}{480.16}\ga^{\rfive}\ga_{\mu}\e_{1}\lb\ga_{\rfive}\la+\frac{i}{32}[(\frac{9}{4}\ga_{\mu}\ga^{\rho}+3\ga^{\rho}\ga_{\mu})\e_{1}\lb\ga_{\rho}\la-\\ \nn
&(\frac{1}{24}\ga_{\mu}\ga^{\rrr}+\frac{1}{6}\ga^{\rrr}\ga_{\mu})\e_{1}\lb\ga_{\rrr}\la+\frac{1}{960}\ga_{\mu}\ga^{\rfive}\e_{1}\lb\ga_{\rfive}\la]+\cdots
,
\ena
where the first term comes from $\hat{F}_5$. 
The contribution of the first term to  $\deps^{5}\dep^{0}\psi_{\mu}$ is
\bee
49/64 i
g_{1}\gf\la^{15}_{d}(\ga_{\mu}\ga^{\sigma})_{ad}\bar{\e_{2}}\ga_{\sigma}\e_{1} .
\ene

The contribution to $\deps^{5}\dep^{0}\psi_{\mu}$ from the term in
equation (\ref{g1}) proportional to $\lb\ga_{\rho}\la$ can be
shown to vanish.
The term proportional to $\lb\ga^{\rrr}\la$ in equation (\ref{g1})
yields the following contribution to  $\deps^{5}\dep^{0}\psi_{\mu}$
\bee
-(9/4+3/2)i
g_{1}\gf\la^{15}_{d}(\ga_{\mu}\ga^{\sigma})_{ad}\bar{\e_{2}}\ga_{\sigma}\e_{1} .
\ene
Using $\ga^{\rfive}\ga^{\mu}\ga_{\rfive}=0$, we can show that the term
proportional to $\lb\ga_{\rfive}\la$ does not contribute. Thus we get
\bee
\deps^{5}\dep^{0}\psi_{\mu a}=-\frac{191}{64}ig_{1}\gf\la^{15}_{d}(\ga_{\mu}\ga^{\sigma})_{ad}\bar{\e_{2}}\ga_{\sigma}\e_{1} .
\ene

Demanding the closure of the supersymmetry algebra on using the equations of motion gives
\bee
ig_{2}+191i g_{1}=\frac{1}{2}\fon . \label{main4}
\ene

In order to derive the above result, one needs the following equation which can be obtained by considering the lowest order supersymmetry transformation on the gravitino,
\bee
[\dep^{0},\deps^{0}]\s_{\mu}=-\frac{i}{64}(\bar{\e_{2}}\ga_{\sigma}\e_{1})\ga^{\sigma}\ga_{\mu}^{\,\,\rho\lambda}\rm{D}_{\rho}\s_{\la}+\cdots ,
\ene
where the ellipsis indicate terms that are not needed in the calculation.
Using equations (\ref{main1}),(\ref{main2}),(\ref{main3}) and
(\ref{main4}) we get:

\bee
\Dft\fof=\left(-182+\frac{15}{4}\right)\fof ,
\ene
which is what is expected from equation (B.16).

The solution for the equation above is given by
${\rm D}^{14}E_{5/2}$ where $E_{5/2}$ is the Eisenstein series
of order $5/2$. The expansion of the Eisenstein series as given in
equation (\ref{eis}) suggests that
the $\gl$ term we have been considering receives correction from
tree-level and two-loops and a series of non-perturbative D-instanton
contributions. This is consistent with the generalized
higher-derivative conjectures in the literature\cite{mbgss,berk}.

\section{Discussion}
In this paper we have concluded the supersymmetry argument initiated
in \cite{mbgss} for the $\gl$ term in the type IIB effective
action. We have shown that the coefficient of this does satisfy the
expected eigen-value equation on the fundamental domain of $SL(2,{\Bbb
Z})$. This gives further evidence for the conjecture given by equation
(\ref{conj}). Generalizing such arguments based on supersymmetry for higher order terms seems to get far more tedious owing to the mixing at other orders of $\alpha'$.

The preceding argument was too crude to distinguish between the three
different contractions in equation (\ref{tensor}). In principle, it
should be possible to obtain a superspace formulation of these terms
which are given by integration over 3/4 of the Grassmann
coordinates. However there is no obvious covariant way of doing
this. One can motivate the structure of the contractions in $\gl$
interaction by reference to the analysis of the $\hg^{4}C^{4}$ term in
\cite{berk}. This starts by considering certain six-dimensional
superstring scattering amplitudes which can be expressed as
topological computations on the hyper-Kahler compactification
manifold\cite{berk2}. The $\gf$ factor in $\gl$ is necessarily a
Lorentz singlet which is not the case in the $\hg^{4}C^{4}$
term. However, it can be argued that the singlet part of $\hg^{4}$
contraction in \cite{berk} is identical to the $\gf$ factor in $\gl$.
As a result it is very suggestive that the $\gf$ factor is given by

\bee
\gf=a\left(T_{1}+15(T_{2}-6T_{3})\right) ,
\ene
where

\bea
T_{1}&=&-G_{\m1\n1\r1}G^{\m1\m2\r1}G_{\m3}^{\,\,\,\,\n1\n2}G_{\m2\n2}^{\,\,\,\,\,\,\,\,\,\,\m3} \\
T_{2}&=&G_{\m1\n1\r1}G^{\m1\m3\n2}G_{\m3\m4}^{\,\,\,\,\,\,\,\,\,\,\,\,\r1}G^{\m4\n1}_{\,\,\,\,\,\,\,\,\,\,\n2} \\
T_{3}&=&G_{\m1\n1\r1}G^{\m1\n1\r1}G_{\m2\n2\r2}G^{\m2\n2\r2} .
\ena
The coefficients in equation (\ref{tensor}) are thus given by $b=-15a, c=-90a$.

It would be gratifying to have a direct computation in string theory
of the D-instanton contributions to the $\gl$ interaction, but this
eems to be problematic at present.
  
\section*{\small{Acknowledgements}}
The author is grateful to Michael B. Green for suggesting the problem and for numerous illuminating discussions. The author thanks Nathan Berkovits for
useful correspondence and Stefano Kovacs for discussions. Extensive use of the package GAMMA for mathematica
written by Ulf Gran has been made. This work has been supported by the Gates Cambridge Trust
and the Perse scholarship of Gonville and Caius College, Cambridge.

\newpage
\begin{appendix}
\section{Relevant formulae in IIB supergravity}

The bosonic fields of the IIB supergravity comprise of the graviton,
the antisymmetric two form with a three-form field strength and the
dilaton. The fermionic fields are the gravitino and the dilatino. Spinors in IIB are complex Weyl spinors. The gravitino $\psi_{\mu}$
and the dilatino $\lambda$ have opposite chiralities, the
supersymmetry parameter has the same chirality as the gravitino. The
conjugate of any spinor is defined by
$\bar{\lambda}=\lambda^{*}\gamma^{0}$. The metric is spacelike and the
gamma matrices are real. We make extensive use of various identities
quoted in \cite{mbgss}. The Fierz identity for ten-dimensional complex
Weyl spinors of the same chirality is:

\bee
\lambda_{1}^{a}\bl^{b}_{2}=-\frac{1}{16}\bl_{2}\gamma_{\mu}\la_{1}\gamma^{\mu}_{ab}+\frac{1}{96}\bl_{2}\gamma_{\mu\nu\rho}\la_{1}\gamma^{\mu\nu\rho}_{ab}-\frac{1}{3840}\bl_{2}\gamma_{\r1...\r5}\la_{1}\gamma^{\r1...\r5}_{ab}
\ene

The bosonic fields which appear are supercovariantized in the
following way.
\bea
\hg_{\mu\nu\rho}&=&G_{\mu\nu\rho}-3\bar{\psi}_{[\mu}\gamma_{\nu\rho]}\lambda-6i\bar{\psi^{*}}_{[\mu}\gamma_{\nu}\psi_{\rho]}
\nonumber \\
\hat{F}_{\r1...r5}&=&F_{\r1...\r5}-5\bar{\psi}_{[\r1}\gamma_{\r2\r3\r4}\psi_{\r5]}-\frac{1}{16}\bar{\lambda}\gamma_{\r1...\r5}\lambda
\ena

The lowest order supersymmetry transformation for the various fields
are given below(we retain only the relevant portions, for the complete
transformations, see \cite{mbgss,sch}). For $\tau$

\bee
\zerd\tau=2\tau_{2}\bar{\epsilon}^{*}\lambda
\ene

The supersymmetry transformation of the zehnbein is given by:

\bee
\delta^{(0)}e^{m}_{\mu}=i(\bar{\epsilon}\gamma^{m}\psi_{\mu}+c.c.)
\ene

The transformation for the dilatino in the fixed $U(1)$ gauge is

\bee
\zerd\la_{a}=..-\frac{i}{24}\gamma^{\mu\nu\rho}\epsilon_{a}\hg_{\mu\nu\rho}+\frac{3}{4}i\lambda_{a}(\bar{\epsilon}\la^{*})-\frac{3}{4}i\la_{a}(\bar{\epsilon^{*}}\la)
\ene
where the last two terms come from the compensating $U(1)$ gauge
transformation.

The gravitino transformation is given by
\bea
\delta^{(0)} \psi_\mu =& D_\mu \epsilon +
{1\over 480}i \gamma^{\rho_1\dots\rho_5}\gamma_\mu \epsilon \hat F_{\rho_1\dots
\rho_5}+ {1\over 96}\left(\gamma_\mu^{\ \nu\rho\lambda}\hat G_{\nu\rho\lambda}
-
9 \gamma^{\rho\lambda} \hat G_{\mu\rho\lambda}\right)\epsilon^*  \\ \nonumber
&-{7\over 16} \left(\gamma_\rho \lambda\, \bar \psi_\mu \gamma^\rho \epsilon^*
- {1\over 1680} \gamma_{\rho_1\dots\rho_5} \lambda \, \bar\psi_\mu
\gamma^{\rho_1\dots\rho_5} \epsilon^* \right)
+ {1\over 32} i \left[\left({9\over 4} \gamma_\mu\gamma^\rho +
3\gamma^\rho \gamma_\mu\right)\epsilon\, \bar \lambda
\gamma_\rho\lambda \right.\\ \nonumber
&\left.   -\left({1\over 24} \gamma_\mu \gamma^{\rho_1\rho_2\rho_3} +
{1\over 6} \gamma^{\rho_1\rho_2\rho_3} \gamma_\mu\right)
\epsilon\,\bar\lambda \gamma_{\rho_1\rho_2\rho_3}\lambda +
{1\over 960} \gamma_\mu\gamma^{\rho_1\dots\rho_5} \epsilon\, \bar
\lambda \gamma_{\rho_1\dots\rho_5} \lambda    \right] + \delta^{(0)}_\Sigma (\psi_\mu),
\ena
where the compensating $U(1)$  transformation is given by
\bee
\delta^{(0)}_\Sigma\psi_\mu
= {1\over 2} i \Sigma = {1\over 4} i\psi_\mu
(\bar\epsilon \lambda^*)  - {1\over 4} i
\psi_\mu (\bar\epsilon^* \lambda)
\ene

\section{Modular covariance and formulae}

The various coefficient functions in the effective action are
($w,\hw$) forms, where $w$ refers to the holomorphic modular weight
and $\hw$ to the anti-holomorphic modular weight. A nonholomorphic
modular form $F^{(w,\hw)}$ transforms as,

\bee
F^{(w,\hw)}\rightarrow F^{(w,\hw)}(c\t+d)^{w}(c\bt+d)^{\hw} \label{F1}
\ene

under the $SL(2,{\Bbb Z})$ transformation taking,

\bee
\t\rightarrow\frac{a\t+b}{c\t+d}
\ene

where $ad-bc=1$. Equation (\ref{F1}) describes a $U(1)$ transformation
when $\hw=-w$. We define the modular covariant derivative

\bee
\cD_{w}=i\left(\pt-i\frac{w}{2\tb}\right)
\ene

This maps $F^{(w,\hw)}$ to $F^{(w,\hw+2)}$. We define

\bee
{\rm D}_{w}\equiv\tb\cD_{w}, \quad {\bar{\rm D}}_{\hw}\equiv\tb\bar{\cD}_{\hw}
\ene

This has the effect

\bee
{\rm D}_{w}F^{(w,\hw)}=F^{(w+1,\hw-1)}, \quad {\bar{\rm
D}}_{\hw}F^{(w,\hw)}=F^{(w-1,\hw+1)}
\ene

The laplacian on the fundamental domain of $SL(2,{\mathbb Z})$ is defined
to be

\bee
\nabla_{0}^{2}=4\tb^{2}\pt\ptb
\ene

when acting on (0,0) forms. The laplacians acting on $(w,-\hw)$ are
defined as

\bea
\nabla_{(-)w}^{2}\equiv
4\rD_{w-1}\brD_{-w}&=&4\tb^{2}\pt\ptb-2iw(\pt+\ptb)-w(w-1)\\
\nabla_{(+)w}^{2}\equiv
4\brD_{-w-1}\rD_{w}&=&4\tb^{2}\pt\ptb-2iw(\pt+\ptb)-w(w+1)
\ena

If $\nabla_{(-)w}^{2}F^{(w,-w)}=\sigma_{w}F^{(w,-w)}$ then

\bee
\nabla^{2}_{(-)w-m}F^{(w-m,-w+m)}=(\sigma_{w}+2mw-m^{2}-m)F^{(w-m,-w+m)}
\ene

and a similar one for $\nabla^{2}_{w-m}$. Using the above relations,
it can be shown \cite{mbgss} that if as in the case of $\alpha'^{3}$,
the function $f^{(12,-12)}$ is an eigenfunction of
$\nabla_{(-)12}^{2}$ satisfying

\bee
\nabla_{(-)12}^{2}f^{(12,-12)}=(-132+\frac{3}{4})f^{(12,-12)} \label{al3}
\ene

then $f^{(0,0)}$ which is the coefficient of the ${\cal R}^{4}$ term
satisfies

\bee
\nabla_{0}^{2}f^{(0,0)}=\frac{3}{4}f^{(0,0)} \label{al31}
\ene

The solution to the above equation is unique if $f^{(0,0)}$ has a
power law behaviour near the boundary of the fundamental domain of
$SL(2,{\mathbb Z})$ which is in agreement with known tree-level and one
loop calculations. If the eigenvalue can be written as $s(s-1)>1/4$ then
the solutions are well known and can be expressed in terms of the
nonholomorphic Eisenstein series

\bee
E_{s}(\t)=\frac{1}{2}\tb^{s}\sum_{(m,n)=1}|m\t+n|^{-2s}
\ene

where ($m,n$) denotes the greatest common divisor of $m$ and $n$.

The asymptotic form of the Eisenstein series for large $\tau_{2}$ can
be found by manipulating the series using a Poisson resummation. The
general formula for $f^{(0,0)}_{s}$ that is the solution for the above
equation for $s$ is

\bee
f^{(0,0)}=a_{s}\tb^{s}+b_{s}\tb^{1-s}+\frac{2\sqrt{\tb}\pi^{s}}{\Gamma(s)}\sum_{(m,n)\neq(0,0)}|\frac{m}{n}|^{s-1/2}K_{s-1/2}(2\pi\tb|mn|)e^{2\pi
mn\tau_{1}} \label{eis}
\ene
where
\bee
a_{s}=2\zeta(2s) \quad \quad
b_{s}=2\sqrt{\pi}\zeta(2s-1)\frac{\Gamma(s-\frac{1}{2})}{\Gamma(s)}
\ene

where $K_{s}(x)$ is the standard modified Bessel function whose
expansion for large $x$ is given by

\bee
K_{r}(x)=(\frac{\pi}{2x})^{\frac{1}{2}}e^{-x}\left[\sum_{n=0}^{\infty}\frac{1}{(2x)^{n})}\frac{\Gamma(r+n+\frac{1}{2})}{\Gamma(r-n+\frac{1}{2})\Gamma(n+1)}\right]
\ene

In the case we are considering, $s=5/2$. Thus the eigenvalue in
equation (\ref{al31}) is $15/4$. The coefficient of $\gl$ is given by
the solution to the equation
\bee
\nabla_{(-)14}^{2}f_{2}^{(14,-14)}=(-196+14+15/4)f_{2}^{(14,-14)}
\ene

\end{appendix}

\end{document}